\documentclass{emulateapj}

\usepackage{lscape}

\newcommand{\kms}{km s$^{-1}$}
\newcommand{\msun}{M$_{\odot}$}
\newcommand{\ntot}{488}
\newcommand{\ngal}{439}
\newcommand{\nhvc}{49}
\def\be{\begin{equation}}
\def\ee{\end{equation}}

\shorttitle{ALFALFA Catalog of the Anti-Virgo Region}
\shortauthors{A. Saintonge et al.}

\begin{document}

\title{The Arecibo Legacy Fast ALFA Survey: V. HI Source Catalog \\
of the Anti-Virgo Region at $\delta=+27^{\circ}$}

\author {Am\'elie Saintonge\altaffilmark{1,2}, 
Riccardo Giovanelli\altaffilmark{1,3}, Martha P. Haynes\altaffilmark{1,3}, 
G. Lyle Hoffman\altaffilmark{4}, \\
Brian R. Kent\altaffilmark{1}, Ann M. Martin\altaffilmark{1}, Sabrina Stierwalt\altaffilmark{1} \&
Noah Brosch\altaffilmark{5}
}

\altaffiltext{1}{Center for Radiophysics and Space Research,
Cornell University, Ithaca, NY 14853. {\it e--mail:} amelie@astro.cornell.edu, 
riccardo@astro.cornell.edu, haynes@astro.cornell.edu, bkent@astro.cornell.edu, 
amartin@astro.cornell.edu, sabrina@astro.cornell.edu}

\altaffiltext{2}{now at: Institute for Theoretical Physics, University of Z\"{u}rich, CH-8057 Z\"{u}rich, Switzerland}

\altaffiltext{3}{National Astronomy and Ionosphere Center, Cornell University,
Ithaca, NY 14853. The National Astronomy and Ionosphere Center is operated
by Cornell University under a cooperative agreement with the National Science
Foundation.}

\altaffiltext{4}{Department of Physics, Lafayette College, Easton, PA 18042. 
{\it e--mail:} hoffmang@lafayette.edu}

\altaffiltext{5}{Wise Observatory and School of Physics and Astronomy, Raymond and Beverly Sackler Faculty of Exact Sciences, Tel Aviv University, Israel. {\it e-mail:} noah@wise.tau.ac.il}

\begin{abstract}
We present a second catalog of HI sources detected in the Arecibo Legacy Fast ALFA Survey. We report \ntot \ detections over 135 deg$^2$, within the region of the sky having $22{\rm h}<\alpha<03{\rm h}$ and $+26^{\circ}<\delta<+28^{\circ}$. We present here the detections that have either (a) $S/N>6.5$, where the reliability of the catalog is better than 95\% or (b) $5.0<S/N<6.5$ and a previously measured redshift that corroborates our detection. Of the \ntot \ objects presented here, \nhvc \ are High Velocity Clouds or clumps thereof with negative heliocentric recession velocities. These clouds are mostly very compact and isolated, while some of them are associated with large features such as Wright's Cloud or the northern extension of the Magellanic Stream. The remaining \ngal \ candidate detections are identified as extragalactic objects and have all been matched with optical counterparts. Five of the six galaxies detected with M$_{HI}<10^{7.5}$M$_{\odot}$ are satellites of either the NGC672/IC1727 nearby galaxy pair or their neighboring dwarf irregular galaxy NGC784. The data of this catalog release include a slice through the Pisces-Perseus foreground void, a large nearby underdensity of galaxies. We report no detections within the void, where our catalog is complete for systems with HI masses of $10^8$M$_{\odot}$. Gas-rich, optically-dark galaxies do not seem to constitute an important void population, and therefore do not suffice at producing a viable solution to the void phenomenon.
\end{abstract}

\section{Introduction}
The Arecibo Legacy Fast ALFA Survey (ALFALFA) is an on-going project conducted at the Arecibo Observatory. Using the 305m telescope and the seven-beam ALFA receiver, the survey aims at covering 7000 deg$^2$ of high galactic latitude sky in order to detect some 20,000 extragalactic HI-bearing objects \citep[]{alfalfa1}. As first illustrated in \citet{alfalfa2}, the survey makes significant improvement over previous blind HI surveys in terms of sensitivity, resolution, and number of detections. Earlier this year, in a first ALFALFA data release, we presented a catalog of 730 HI detections for the northern Virgo Cluster region \citep{alfalfa3}. Of these detections, only $5\%$ are present in the northern HIPASS catalog \citep{wong} over the same area of the sky, and $69\%$ of the cataloged sources are new HI detections. Clearly, the contribution of ALFALFA is significant.

The survey was designed to cover a wide variety of cosmic environments. The ``Spring'' ALFALFA survey covers the right ascension range from 7.5 to 16.5 hours, which includes several nearby overdensities such as the Virgo Cluster, the Coma Cluster and the Leo Group, while the ``Fall'' survey maps the region from 22 to 3 hours of right ascension over most of the declination range accessible from Arecibo ($\delta=0^{\circ}-36^{\circ}$). This region includes the Pisces-Perseus Supercluster and also a large nearby cosmic void in its foreground. We present in this paper a first catalog of galaxies from this ``Fall'' survey. During the first season of ALFALFA observing, all observations necessary to obtain full coverage over a $2^{\circ}$ wide strip were completed. This specific region of sky, centered on $\delta=+27^{\circ}$ and stretching from $\alpha=22$h to $\alpha=3$h, was chosen to begin the survey observations because it was already targeted during the ALFALFA precursor observations \citep{alfalfa2}, it goes through the very nearby galaxy group NGC672, it covers the southernmost portion of the Pisces-Perseus Supercluster ridge, and surveys a large nearby void.

The scientific goals and applications of ALFALFA are very broad and diverse \citep{alfalfa1}; an important one  is the determination of the HI mass function (HIMF), and its dependence on environment. 
Several blind HI surveys have been used to address this question, both using single dish telescopes \citep[e.g.][]{briggs,rosenberg,hipass,hijass} and synthesis arrays \citep[e.g.][]{weinberg,deblok}. While there is no consensus yet on the slope of the faint end of the HIMF \citep[for a recent compilation of the different results, see][]{springob05}, the different surveys measure an abundance of low-mass HI systems that is far less than the number of low-mass dark matter halos produced by numerical simulations \citep[e.g.][]{klypin99}. With its sensitivity and large sky coverage, ALFALFA should detect hundreds of galaxies with HI masses less than $10^{7.5}$M$_{\sun}$, compared to the handful contained in other surveys, and should therefore help determine with high confidence the abundance of low-mass HI systems in the local Universe and put constraints on the ``missing satellites'' problem.

There is another related problem when it comes to comparing observations of galaxies with the predicted abundances from simulations, which can also be addressed by ALFALFA. This problem concerns the abundance of low-mass 
systems in underdense regions of space and has been coined as the ``void phenomenon'' \citep{peebles01}. Using numerical simulations, \citet{gott03} find that voids are filled with dark matter filaments, just like denser environments, except that the amplitudes of the perturbations are much smaller. They predict that a void with a diameter of 20 $h^{-1}$ Mpc should contain almost 1000 dark matter haloes with masses of $10^9 h^{-1}$ M$_{\sun}$ and about 50 with masses of $10^{10} h^{-1}$ M$_{\sun}$. Unless there are mechanisms at play that would inhibit the formation of galaxies in underdense regions of space, there should therefore be a significant population of dwarf galaxies in voids. Such a population has so far not been observed. \citet{hoyle05} have built a luminosity function of void galaxies using data from the Sloan Digital Sky Survey. They do not find any significant dependence of the slope of the faint end of the luminosity function on environment. A steeper slope in the voids would have revealed the existence of a more important population of dwarfs.

If we accept the predictions of the numerical simulations, there are two main avenues to explain the discrepancy with the results of the optical surveys: (1) the processes that are required to trigger the star formation process in galaxies are not available in voids, or (2) there is no material out of which these stars could be formed.  While the second scenario implies that some physical processes have suppressed star formation in dwarf galaxies by removing their reservoirs of cool gas, according to the first scenario the voids should contain a large number of halos that are filled with HI but devoid of stars, the so-called ``dark galaxies''. So far, such a population has not been observed, though a few candidates have been reported \citep{schneider83,gh89,kilborn00,ryder01,minchin}. After further investigation, these have however all been found to be either associated with or the companions of a faint optical counterpart, or part of the tidal features of neighboring galaxies \citep[][Haynes et al. 2007 in preparation]{schneider89,chengalur95,doyle05}. The largest blind HI survey to date, the HI Parkes All Sky Survey \citep[HIPASS,][]{hipass} has failed to find any truly dark galaxies. With a sensitivity greater than that of HIPASS by a factor of eight, ALFALFA promises to either find a population of dark galaxies if they exist, or at least to put a much stronger constraint on their abundance. In this paper, we provide a new catalog of HI detections from ALFALFA which provides first insight on a slice through the Pisces-Perseus foreground void and allows us to investigate in a preliminary manner what ALFALFA may be telling us about the void phenomenon.

The rest of the paper is organized as follows. In \S 2, we describe the large-scale structure present in the sky area studied, give an overview of the ALFALFA observing and data reduction processes and present a catalog of \ntot \ HI detections. In \S 3, we present the properties of the sample and a description of the High Velocity Cloud population. Finally, in \S 4 we discuss the implications of the data for the void problem. A value for the Hubble constant of 70 \kms \ Mpc$^{-1}$ has been assumed throughout the paper.

\section{Data}

\subsection{Sky Coverage and Large Scale Structure \label{lss}}
We present data for a $2^{\circ}$-wide strip of the ALFALFA survey, covering a total of 135 deg$^2$. The strip is centered at $+27^{\circ}$ declination and covers the right ascension range from $22^{\rm h}00^{\rm m}$ to $03^{\rm h}04^{\rm m}$.  Observations were made between August and November 2005 and required just over 100 hours of telescope time.
The declination strip to be first observed as part of the ALFALFA ``Fall'' survey and to be presented in this data release was selected both for technical and scientific reasons. The part of this strip between $00^{\rm h}30^{\rm m}$ to $02^{\rm h}30^{\rm m}$ was surveyed in a shared-risk mode during Fall 2004 as part of the ALFALFA precursor observations \citep{alfalfa2}. This specific region was selected then, because with a moderate zenith angle of $\sim10^{\circ}$, both sensitivity and pointing accuracy of the telescope were maximized.  For these same reasons, to complete the coverage over the full RA range, and to assess the performance of the full ALFALFA infrastructure that was perfected during these precursor observations, the strip was retargeted when the full survey was initiated.

In order to get some early science results, this declination range was also chosen because it allows the study of interesting features of the local Universe, the nearby galaxy group NGC672/IC1727 for example. It also allows us to start studying several important features of the large scale structure. The Pisces-Perseus Supercluster (PPS) is one of the most prominent nearby extragalactic features. The main ridge of the cluster stretches for at least $50h^{-1}$Mpc, has a very large axial ratio with a width in the plane of the sky of only $5-10h^{-1}$Mpc \citep{wegner93}, and stretches almost perpendicularly to the line of sight. The southernmost part of the ridge crosses the sky area studied here. In the late-80's, a large project was undertaken to obtain redshifts for a large number of spiral galaxies in the PPS region. HI 21cm spectra were obtained using the Arecibo Telescope \citep{gh85,giovanelli86,gh89PPS} for these galaxies selected from the {\it Catalog of Galaxies and Clusters of Galaxies} 
\citep[CGCG][]{cgcg}.
Redshift measurements show a significant enhancement in the distribution of galaxies at $cz\sim5000$ \kms, and a large underdensity in the foreground over the entire PPS area they surveyed.  Radio synthesis observations were also conducted over a set of twelve selected field in the direction of the PPS by \citet{szomoru94}, These authors concluded that the properties of the galaxies in their HI-selected sample did not differ substantially from those of late-type galaxies found in optical surveys. They also report that while 16 objects were detected in the supercluster region, none were detected over the same volume and down to the same mass sensitivity limit of $\sim5\times10^7$\msun \ in the foreground void.

This void in the foreground of PPS was first identified by \citet{hg86}. Using a total of nearly 3000 galaxies with redshifts, they recognized the filamentary structure in the distribution of galaxies in and around PPS, the presence of narrow structures appearing to connect PPS and the Local Supercluster, but mostly the void nature of a major portion of the foreground between PPS an the Local Supercluster. That void is most prominent at declinations between $+30^{\circ}$ and $+50^{\circ}$, where most of the volume between 22h and 3h and extending from $cz\sim1000$\kms \ to $cz\sim4000$\kms \ is empty. At lower declinations, there also appears to be underdense regions at $cz\lesssim2500$\kms \ \citep[see Fig.2 of ][]{hg86}.

\subsection{Observations and Data Reduction \label{data}}
ALFALFA observations are conducted at the Arecibo telescope and the 7-feed ALFA receiver.
As described elsewhere \citep{alfalfa1,alfalfa2,alfalfa3}, data acquisition for ALFALFA is done in a fixed azimuth drift scan mode. Each sky tile is covered in two partly-overlapping drifts at two different epochs in order to achieve better than Nyquist sampling. Moreover, since the two passes are separated in time by several months, comparing the data from the two epochs can help in ruling out spurious detections. When data acquisition is completed over a given region of the sky, the individual drift scans are assembled and regridded to form three-dimensional data cubes or ``grids''. These grids are $2.4^{\circ} \times 2.4^{\circ}$ in size with 1\arcmin \ sampling, and have 1024 channels along the spectral axis. Since the velocity resolution is $\sim5.3$ \kms (before Hanning smoothing), in order to cover the full ALFALFA velocity range which extends from -2000 to 18,000 \kms, each grid is separated in four redshift subgrids, respectively covering the ranges: -2000 to 3300\ \kms, 2500 to 7900\ \kms, 7200 to 12800\ \kms, and 12100 to 17900\ \kms. We present here the detections made in 38 such grids. The full data reduction and gridding process will be presented in detail in Giovanelli et al. (2008, in preparation).

\begin{figure}[h]
\epsscale{0.9}
\plotone{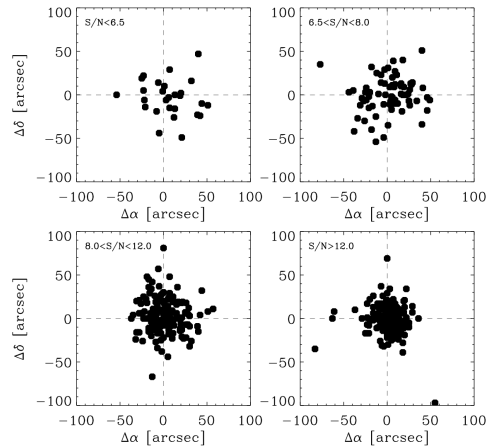}
\caption{Pointing offsets between measured HI coordinates and the coordinates of the assigned optical counterparts, in four different $S/N$ bins. The median pointing error is 25\arcsec, 25\arcsec, 21\arcsec \ and 14\arcsec, respectively for the four plots from top-left to bottom-right. \label{pointing}}
\end{figure}

Signals were identified using the ALFALFA automated signal extraction tool, that creates a catalog of candidate detections by convolving in Fourier space the data with templates build from a combination of Hermite functions \citep{alfalfa4}. Galactic emission is fitted and subtracted to allow for the detection of objects with very low recession velocities. In cases where an extragalactic signal straddles the spectral window contaminated by Galactic emission, an interpolation tool is used before parameters such as width and flux can be measured.  Each of the candidate detections was then examined, measured, and cross-correlated with DSS2, NED, and the AGC (``Arecibo General Catalog'', a database of galaxies maintained by M.P.H and R.G.). No data is available from the {\it Sloan Digital Sky Survey} over this part of the sky. Optical counterparts were at this point assigned to the HI detections, mostly unambiguously due to the pointing accuracy and good spatial resolution of the survey. Figure \ref{pointing} shows the pointing offsets between measured HI coordinates and the positions of the optical counterparts, in four different $S/N$ bins. A correction has been applied to account for the systematic pointing errors of the telescope by comparing the observed positions of continuum sources with those given in the NVSS catalog \citep{nvss}. After this correction, the median pointing offsets for all detections in the catalog is 19\arcsec, while the error is reduced to 14\arcsec \ for detections with $S/N>12$. The outliers seen in Figure \ref{pointing} are detections made in the vicinity of other systems, where the centroiding process is complicated by this external contamination. In all cases, a note to that effect is made in \S \ref{cat}. Given the $3.3^{\prime} \times3.8^{\prime}$ size of the beam and the spatial resolution of previous large blind HI surveys, the accuracy of the coordinates of the HI detections in the gridded ALFALFA data cubes is excellent.

\subsection{Catalog Presentation \label{cat}}

Three different types of detections are presented here. We include in the catalog any detection with $S/N>6.5$ made in regions of the data cubes that are not affected by radio frequency interference and where the survey coverage is good. ALFALFA catalogs are better than 95\% reliable above this threshold \citep{alfalfa4}, and such detections do not require any corroborating follow-up observations (code 1 in Tab. \ref{params}). The second class of objects presented in this catalog has $5.0<S/N<6.5$, but a previously measured optical redshift with which the HI measurement agrees (code 2 in Tab. \ref{params}). We consider that this prior information significantly increases the reliability of these detections, to the point where they can also be included in this catalog without further observations. Finally, we also include High Velocity Clouds (HVCs), which in this part of the sky are all found to have large negative recession velocities (code 9). These HVCs are discussed in more detail in \S \ref{hvc}. Figure 2 of \citet{alfalfa3} shows examples of HI line profiles corresponding to the different code assignments.

We present in Table \ref{params} the measured parameters for \ntot \ detections, \ngal \ of which are associated with extragalactic objects, while the remaining \nhvc \ are HVC features. The contents of the different columns are:
\begin{itemize}
\item Col. 1: an entry number for this catalog

\item Col. 2: the source number in the Arecibo General Catalog, a private database
        of extragalactic objects maintained by M.P.H. and R.G. The AGC entry normally
        corresponds to an optical counterpart except in the cases of HI sources 
        which cannot be associated with an optical object with any high
        degree of probability.

\item Cols. 3 \& 4: center (J2000) of the HI source, after correction for systematic 
	telescope pointing errors, which are on the order of 20\arcsec.
	The accuracy of the HI positions depends on source strength. On average,
	the positional accuracy is about 19\arcsec. See \S \ref{data} for
	details.

\item Cols. 5 \& 6: center (J2000) of the optical galaxy found to provide a reasonable
	optical counterpart. This position has been checked for each listed
	object and assessed using tools provided through the {\it SkyView} website.
	Quality of centroids is estimated to be 2\arcsec~ or better. The assessment
	of optical identification is based on spatial proximity, optical
	redshift (if available), morphology and color. For sources with no discernible optical
	counterpart and those for which such assignment is ambiguous, due to 
	the presence of more than one equally
	possible optical counterpart, no optical position is listed. The
	latter set includes HVCs. For objects with more than
	one possible candidate as an optical counterpart but such that one
	of the candidates is significantly more preferable than the others,
	an optical identification is made; however, a comment on
	the possible ambiguity is added in the notes to this table, as
	alerted by an asterisk in Col. 14.

\item Col. 7: heliocentric velocity of the HI source, $cz_{\odot}$, measured 
	as the midpoint 
	between the channels at which the flux density drops to 50\% of 
	each of the two peaks (or of one, if only one is present) at each
	side of the spectral feature. Units are \kms. The error on $cz_\odot$
	to be adopted is half the error on the width, tabulated in Col. 8.

\item Col. 8: velocity width of the source line profile, $W50$, measured at the 50\%
	level of each of the two peaks, as described for Col. 7. This value 
	is corrected for instrumental broadening. No corrections due to
	turbulent motions, disk inclination or cosmological effects are
	applied. In parenthesis we show the estimated error on the velocity width, $\epsilon_w$, in \kms.
	This error is the sum in quadrature of two components: the first is a
	statistical error, principally dependent on the S/N
	ratio of the feature measured; the second is a systematic error
	associated with the subjective guess with which the observer estimates 
	the spectral boundaries of the feature: maximum and minimum guesses of 
	the spectral extent of the feature are flagged and the
	ratio of those values is used to estimate systematic errors on the
	width, the velocity and the flux integral. In the majority of cases,
	the systematic error is significantly smaller than the statistical
	error; thus the former is ignored.

\item Col. 9: integrated flux density of the source, $F_c$, in Jy \kms . This 
	is measured on the integrated spectrum, obtained by
	spatially integrating the source image over a solid angle of at
        least $7$\arcmin $\times 7$\arcmin ~and dividing by the sum of the survey beam
	values over the same set of image pixels \citep[see][]{shostak}. 
	Estimates of integrated fluxes for very extended sources with
	significant angular asymmetries can be misestimated by our 
	algorithm, which is optimized for measuring sources comparable with
	or smaller than the survey beam. A special catalog with parameters
	of extended sources will be produced after completion of the survey. 
	The issue is especially severe for extended High Velocity Clouds 
	that exceed in size that of the ALFALFA data cubes. In these specific 
	cases, only the flux in the knots of emission is measured. In general, this meant 
	applying the same 
	kind of $S/N$ selection threshold as for the extragalactic signals, with 
	the exception of the southern extension of Wright's cloud, where only a 
	selection of the brightest knots was measured to trace the structure.  
	See Column 14 
	and the corresponding comments for individual objects. 
	The estimated uncertainty of the integrated flux density, in Jy \kms, 
	is given in parenthesis. Uncertainties
	associated with the quality of the baseline fitting are not included;
	an analysis of that contribution to the error will be presented
	elsewhere for the full survey. See description of Col. 8 for the
	contribution of a possible systematic measurement error.

\item Col. 10: signal--to--noise ratio S/N of the detection, estimated as 
        \begin{equation}
	S/N=\left ( \frac{1000F_c}{W50} \right ) \frac{w_{smo}^{1/2}}{\sigma_{rms}}
	\label{eqsn}
	\end{equation}
        where $F_c$ is the integrated flux density in Jy \kms, as listed in Col. 9,
        the ratio $1000 F_c/W50$ is the mean flux across the feature in mJy,
        $w_{smo}$ is either $W50/(2\times 10)$ for $W50<400$ \kms \ or
        $400/(2\times 10)=20$ for $W50 \geq 400$ \kms [$w_{smo}$ is a
        smoothing width expressed as the number of spectral resolution
        bins of 10 \kms \ bridging half of the signal width], and $\sigma_{rms}$
        is the r.m.s noise figure across the spectrum measured in mJy at 10
	\kms \ resolution, as tabulated in Col. 11. In a similar analysis \citep{alfalfa2},  
        we adopted a maximum smoothing 
	width $W50/20=10$, but this expression was revised to the current one, presented in
	\citet{alfalfa3}. The value of the smoothing width could be
	interpreted as an indication of the degree to which spectral smoothing 
	aids in the visual detection of broad signals, against broad--band 
	spectral instabilities. The ALFALFA
	data quality appears to warrant a more optimistic adoption of
	the smoothing width than previously anticipated. 

\item Col. 11: noise figure of the spatially integrated spectral profile, $\sigma_{rms}$,
	in mJy. The noise figure as tabulated is the r.m.s. as measured over the signal-- and
	rfi-free portions of the spectrum, after Hanning smoothing to a spectral
	resolution of 10 \kms.

\item Col. 12: adopted distance in Mpc, $D_{Mpc}$. For objects with $cz_{\odot}>6000$ \kms, 
	the distance is simply  $cz_{cmb}/H_\circ$; $cz_{cmb}$ is the recession velocity
	measured in the Cosmic Microwave Background reference frame and $H_\circ$ is
	the Hubble constant,  for which we use a value of 70 \kms Mpc$^{-1}$.
	For objects of lower $cz_{cmb}$, we use the peculiar velocity model for the local
	universe of \citet{mastersth}, which is based on data from the SFI++ catalog of 
	galaxies \citep{sfi++}. The transition from one model to the other is selected to be at $cz_{\odot}=6000$ \kms, 
	since at this distance errors in the flow model velocities are comparable or larger than  
	those coming from assuming a Hubble flow.
	
\item Col. 13: logarithm in base 10 of the HI mass, in solar units. That parameter is 
	obtained by using the expression $M_{HI}=2.356\times 10^5 D_{Mpc}^2 F_c$. 

\item Col. 14: object code, defined as follows: 
	
	Code 1 refers to sources 
	of S/N and general qualities that make it a reliable detection.
	By ``general qualities'' we mean that, in addition to an approximate
	S/N threshold of 6.5, the signal should  exhibit a good match between
	the two independent polarizations and a spatial extent consistent
	with expectations given the telescope beam characteristics. Thus, some
	candidate detections with $S/N>6.5$ have been excluded on grounds
	of polarization mismatch, spectral vicinity to rfi features or peculiar
	spatial properties. Likewise, some features with $6.0<S/N<6.5$ are included
	as reliable detections, due to optimal overall characteristics of
	the feature, such as a well defined spatial extent, a broad velocity width, and an 
	obvious association with an optical counterpart with previously unknown redshift.
	 The S/N threshold for acceptance of a reliable detection
	candidate is thus soft. In a preliminary fashion, we estimate that
	these detection candidates with $S/N<6.5$ and code 1 in Table \ref{params} are reliable, i.e. they
	will be confirmed in follow--up observations in better than 95\% of
	cases  \citep{alfalfa4}. Follow-up observations planned for 2007 will 
	set this estimate on stronger statistical grounds.

	Code 2 refers to sources of low S/N ($<$ 6.5), which would  
        ordinarily not be considered
	reliable detections by the criteria set for code 1. However, those
	HI candidate sources are matched with optical counterparts with known
	optical redshifts which match those measured in the HI line. These
	candidate sources, albeit ``detected'' by our signal finding algorithm,
	are accepted as likely counterparts only because of the existence of
	previously available, corroborating optical spectroscopy. We refer to
	these sources as ``priors''. We include them in our catalog because
	they are very likely to be real. There are 31 such sources in the present catalog.

	Code 9 refers to objects assumed to be HVCs; no
	estimate of their distances is made.

 	Notes flag. An asterisk in this column indicates a comment is included
	for this source in the text below.
\end{itemize}

Only the first few entries of Table \ref{params} are listed in the printed version of this
paper. The full content of Table \ref{params} is accessible through the electronic version
of the paper but also through our public digital 
archive site\footnotemark \footnotetext{http://arecibo.tc.cornell.edu/hiarchive/alfalfa/}. 
The comments for those sources marked with an asterisk in Column 14 are given here:\\
\\
\footnotesize
2-  4: compact HVC; one of two nearby knots (the other is AGC102578, another HVC from this catalog) \\
2-  5: compact HVC; one of two nearby knots (the other is AGC102576, another HVC from this catalog) \\
2- 34: just south of AGC100101 (001454.8+261953), may be interacting (this signal may be contaminated by the other) \\
2- 36: close companion of AGC102245 (001536.6+271426); blend: HI signals difficult to untangle \\
2- 40: parameters uncertain, signal merging with strong rfi \\
2- 42: possible alternative opt.id with 001805.8+265933, but farther from HI center and LSB \\
2- 47: possible alternative opt.id with 002204.3+270221, 002211.6+270115 but farther from HI center \\
2- 49: possible alternative opt.id with 002255.0+260751 \\
2- 53: opt.id ambiguous; other possible counterpart at 002930.1+271056 but farther from HI center \\
2- 59: compact HVC; near several other compact clouds \\
2- 62: HVC; close in frequency to Galactic HI, but appears distinct \\
2- 65: compact HVC; near several other compact clouds \\
2- 69: possible alternative opt.id with 003928.9+260506, but farther from HI center \\
2- 71: HVC; part of a stream of clouds \\
2- 72: HVC; part of a stream of clouds \\
2- 73: HVC; in a complex of compact clouds \\
2- 76: HVC; part of a stream of clouds \\
2- 78: HVC; in a complex of compact clouds \\
2- 88: HVC; one of the knots in a complex that extends over 1.5 degrees \\
2- 91: HVC; one of the knots in a complex that extends over 1.5 degrees \\
2- 95: HVC; one of the knots in a complex that extends over 1.5 degrees \\
2-109: compact HVC; in the vicinity of other clouds \\
2-111: HVC; part of the Western end of Wright's Cloud \\
2-112: HVC; part of the Western end of Wright's Cloud \\
2-117: HVC; part of the Western end of Wright's Cloud \\
2-118: HVC; compact cloud, West of the Southern extension of Wright's Cloud \\
2-120: HVC; one knot of a small cloud, just West of the Southern extension of Wright's Cloud \\
2-122: HVC; one knot of a small cloud, just West of the Southern extension of Wright's Cloud \\
2-123: HVC; one knot of a small cloud, just West of the Southern extension of Wright's Cloud \\
2-124: possible alternative opt.id with 010312.9+263826, but farther from HI center \\
2-129: HVC; part of the southern extension of Wright's Cloud \\
2-130: HVC; part of the southern extension of Wright's Cloud \\
2-132: HVC; part of the southern extension of Wright's Cloud \\
2-133: HVC; part of the southern extension of Wright's Cloud \\
2-134: HVC; part of the southern extension of Wright's Cloud \\
2-135: HVC; part of the southern extension of Wright's Cloud \\
2-136: possible alternative opt.id with 011257.2+273801 \\
2-139: compact HVC; in the vicinity of other clouds \\
2-144: compact HVC; interesting morphology, a string of knots \\
2-145: HVC; close in frequency to Galactic HI, yet seems to be spectrally distinct \\
2-147: HI signal most likely a blend of AGC110263 (012240.8+265206) and AGC110264 (012243.1+265200) (assigned to the latter due to better match with optical velocity). Signal also merges with strong rfi so parameters are uncertain \\
2-149: possible alternative opt.id with 012608.7+275819, close companion \\
2-151: opt.id ambiguous; possible alternative opt.id with 012946.1+272221 (IRAS sources, cz=12566km/s) \\
2-158: distance estimated at 7.2 Mpc based on membership of the NGC672 group \\
2-160: distance estimated at 7.2 Mpc based on membership of the NGC672 group \citep{hucht00} \\
2-164: distance estimated at 7.2 Mpc based on membership of the NGC672 group \citep{hucht00} \\
2-166: distance estimated at 7.2 Mpc based on membership of the NGC672 group \citep{hucht00} \\
2-168: parameters uncertain, signal partly blended with NGC672. Primary distance measurement of 7.2 Mpc \citep{karach04}. \\
2-169: parameters uncertain, signal partly blended with IC1727. Primary distance measurement of 7.2 Mpc \citep{karach04}. \\
2-170: opt.id ambiguous; possible counterparts at AGC110534 (014835.3+273324), AGC110535 (014835.2+273253); HI signal probably a blend of these two very nearby spirals. Opt.id made based on best coordinate match and later morphology. \\
2-180: primary distance measurement of 4.7 Mpc \citep{karach03} \\
2-189: opt.id ambiguous: possible alternative opt.id with AGC112531 (015954.1+262401,no previous cz) \\
2-191: parameters uncertain, signal may be partly blended with UGC1507 \\
2-194: possible lsb o.c. close HI coords, but larger galaxy 1.24 arcmin away (2032651+264259) \\
2-207: possible alternative opt.id with 021311.0+274639, very lsb system \\
2-208: parameters uncertain, signal possibly blended with AGC122188 (021307.3+274649) \\
2-216: HVC; compact cloud but close to a larger, extended feature (AGC122816, another HVC from this catalog) \\
2-219: HVC; compact cloud but close to a larger, extended feature (AGC122816, another HVC from this catalog) \\
2-228: HVC; extended cloud, close in frequency to Galactic HI \\
2-229: previous optical redshift measurement of cz=10045 \kms, but possible alternative opt.id with 022447.3+260135 \\
2-230: possible alternative opt.id with 022518.6+270235 \\
2-258: possible alternative opt.id with 023137.2+260927, but much less likely \\
2-272: several possible alternative opt.id, but candidate selected on previous cz information and spatial proximity \\
2-289: possible alternative opt.id with 025906.9+271143, irregular lsb galaxy farther from the HI center \\
2-309: HVC; end of the extension of a cloud located 1 degree away (AGC321292, another HVC from this catalog) \\
2-311: HVC; part of the extension of a cloud (AGC321292, another HVC from this catalog) that extends over 1 degree \\
2-312: possible alternative opt.id with 222120.9+274943, edge-on disc \\
2-314: compact HVC; separated in velocity by about 100 km/s from other nearby clouds \\
2-319: HVC; in the vicinity of other clouds \\
2-320: HVC; in the vicinity of other clouds \\
2-323: HVC; in the vicinity of other clouds \\
2-324: HVC; in the vicinity of other clouds \\
2-327: HI signal is most likely a blend between AGC12191 (224840.9+273638) and AGC12193 (224844.2+273500). Opt.id made with 12193 which is closer to the HI coordinates \\
2-330: HVC; faint, in the vicinity of other clouds \\
2-331: HVC; faint, in the vicinity of other clouds \\
2-333: HVC; faint, in the vicinity of other clouds \\
2-346: possible alternative opt.id with small lsb galaxy closer to HI center at 230012.6+260512 \\
2-364: opt.id ambiguous; possible counterparts at 231717.3+275839, 231710.2+275858 \\
2-370: parameters uncertain, HI signal partly blended with AGC12514 (231955.3+260041) \\
2-377: parameters uncertain, signal partly blended with AGC12546 (232141.2+270512) \\
2-378: parameters uncertain, signal partly blended with AGC12543 (232133.5+270705) \\
2-379: possible alternative opt.id with 232136.3+275216.8 but farther from HI center \\
2-380: parameters uncertain, signal partly blended with AGC12546 (232141.2+270512) \\
2-382: poor baseline, galaxy very near 1.3 Jy continuum source \\
2-399: parameters uncertain, signal blended with AGC12626 (232915.2+262245). HI seen to extend from AGC12626 to AGC333204 (232920.2+261741), most likely an interacting pair. \\
2-404: possible alternative opt.id with 233039.8+262145 \\
2-413: parameters uncertain, signal blending with strong rfi \\
2-414: HVC; one of two knots in a small clumpy cloud \\
2-417: HVC; one of two knots in a small clumpy cloud \\
2-421: parameters uncertain, signal blending with strong rfi \\
2-423: possible alternative opt.id with 233705.4+262541, lower surface brightness but closer from HI detection center \\
2-424: possible alternative opt.id with 233759.6+264810 \\
2-431: HVC; lower velocity than other clouds, merging with Galactic emission \\
2-437: possible alternative opt.id with AGC330782 (234244.7+275056), but 1.9arcmin away from HI center. Surprisingly bright HI given the morphology of the galaxy. \\
2-442: possible alternative opt.id with AGC331304 (234323.0+265517) \\
2-443: very rich field, opt.id ambiguous \\
2-450: HVC; faint cloud, near brighter object AGC333230 \\
2-451: parameters uncertain, signal blending with strong rfi \\
2-452: possible alternative opt.id with 234657.7+272012 but farther from HI center \\
2-454: parameters uncertain, signal blending with strong rfi \\
2-458: HVC; patchy, fairly bright cloud, with two fainter clouds nearby \\
2-459: parameters possibly uncertain, signal blending with strong rfi \\
2-466: possible alternative opt.id with 235044.5+271642, a very lsb system \\
2-470: possible alternative opt.id with 235227.1+270901 \\
\normalsize

\section{Results}

\subsection{Overall Properties of the Detections}

We first compare our detections with previous measurements. Part of the sky area studied here was already covered during the ALFALFA precursor observations and a first catalog was presented in \citet{alfalfa2}. That catalog contained 166 detections, 143 of which are within the right ascension and declination range studied here. There are 9 of these 143 detections that are not recovered with sufficient $S/N$ in the full ALFALFA survey, that is generally with $S/N>6.5$. These were all cases where the detection in the precursor data was very marginal and had to be validated by follow-up higher sensitivity pointed observations using the single pixel L-wide receiver at Arecibo \citep{alfalfa2}. On the other hand, the full survey allows for the detection of about 40\% more galaxies over the same sky area, and that even before any corroborating follow-up observations are conducted on the full survey catalog.

A careful reader will notice small discrepancies in the distances calculated in \citet{alfalfa2} and in the present paper. The first study used the peculiar velocity model of \citet{tonry00}, which is based on 300 early-type galaxies with surface brightness fluctuations measurements. In this study, we used the peculiar velocity model of \citet{mastersth}, which uses a much larger sample of galaxies from the SFI++ catalog of galaxies \citep{sfi++}, and as such provides better constraints on distances. These new distances therefore supersede those previously published.

It is also interesting to compare the ALFALFA detections to previous HI and optical measurements. Of the \ngal \ detections presented in Table \ref{params} that have $cz_{\odot}>0$, 274 ($62\%$) are new HI detections and 261 ($59\%$) are new redshifts.  Even over this area that was heavily targeted for HI measurements in the past, more than $60\%$ of the objects found by ALFALFA were never detected in HI before. This result, which is similar to that found by \citet{alfalfa3} in the Virgo Cluster region, shows how optically targeted surveys fail to select the majority of HI-rich objects. A sensitive blind HI survey such as ALFALFA will therefore significantly contribute to our understanding of the distribution of HI in the local Universe.

\begin{figure}[h]
\plotone{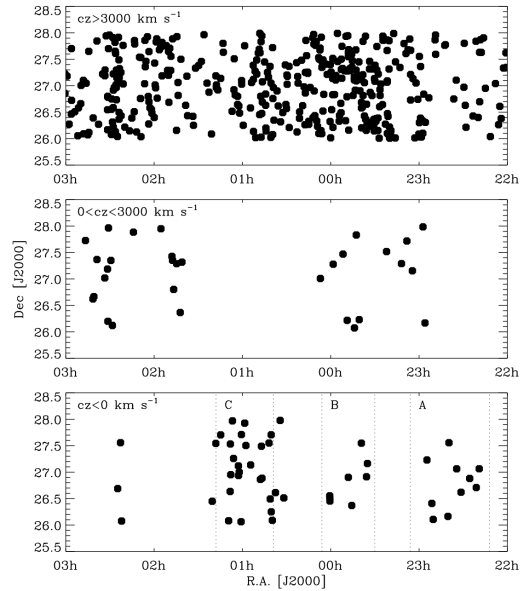}
\caption{Distribution of all the sources on the sky, in three different recessional velocity bins. The bottom panel represents the HVCs and the three regions labeled A, B and C are studied in more detail and contour plots are shown as Figures \ref{hvc22h}, \ref{hvc23h} and \ref{hvc01h}.  \label{sky}}
\end{figure}

In Figure \ref{sky}, we show the distribution on the sky of the detections listed in Table \ref{params}, in three different velocity bins. Some of the large scale structure is already visible in this plot. The overdensity of galaxies in the top panel detected between 23h and 00h of R.A. is associated with a supercluster filament containing the galaxy cluster Abell 2634. The absence of galaxies around 01h with $cz<3000$ \kms \ (middle panel) corresponds to the location of part of a large underdense region, the Pisces-Perseus foreground void. The overall density of extragalactic detections is 3.3 objects per square degree, while that number falls to 0.2 for the objects found within 3000 \kms. As expected, the detection rate is lower than the average 5.4 objects per square degree found in \citep{alfalfa3} where the survey area crosses the supergalactic plane and the northern part of the Virgo cluster. The detections in the bottom panel are HVCs, which mostly have $cz\lesssim-300$\kms. A more detailed description of those is made in \S \ref{hvc}.

\begin{figure}[h]
\plotone{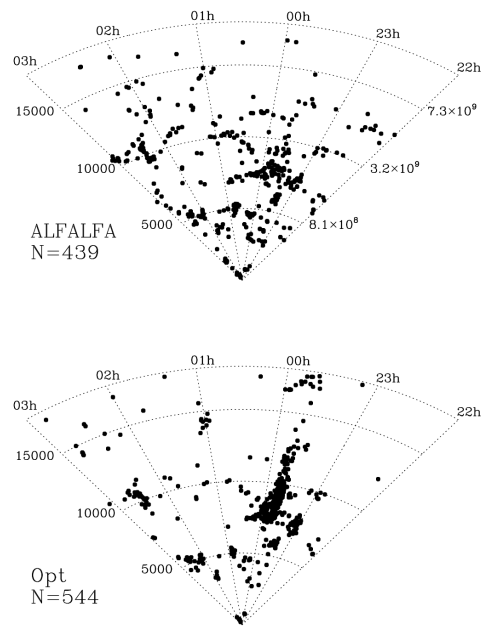}
\caption{Distribution of all the sources in the $2^{\circ}$ declination strip. The top panel shows all the ALFALFA detections presented here, while the bottom panel represents the galaxies with measured optical redshifts in the same volume of space.  In each case, the number of plotted galaxies is reported.  Along the right axis of the top panel, we give the mass limit in solar units for a $7\sigma$ detection of a galaxy with a width $W50$ of 100 \kms \ at the redshift indicated on the left axis.  Note that due to rfi, ALFALFA is blind to cosmic emission between about 15000 and 16000 \kms.\label{cone_plot}}
\end{figure}

The redshift distribution of the extragalactic detections is presented in Figure \ref{cone_plot} as a cone diagram. The top panel is the ALFALFA HI detections, while the bottom panel shows the galaxies with optical redshifts over the same area of the sky.  The optical redshifts are taken from the AGC, which in this part of the sky come mostly from RC3 \citep{RC3} or NED.  Abell 2634 can again be clearly seen, especially in the optical data which includes the data set of \citet{scodeggio95}. The local void is seen again centered on 01h and extending from $cz_{\odot}\sim1000$\kms out to $cz_{\odot}\sim2500$\kms, in both the HI and optical data.  The mass sensitivity for a galaxy with a width of 100 \kms \ detected with $S/N=7$ is also indicated. The lack of ALFALFA detections with $15000\lesssim cz \lesssim 16000$ is due to strong rfi coming from a San Juan airport radar which operates at 1350 MHz. 

\begin{figure}[h]
\plotone{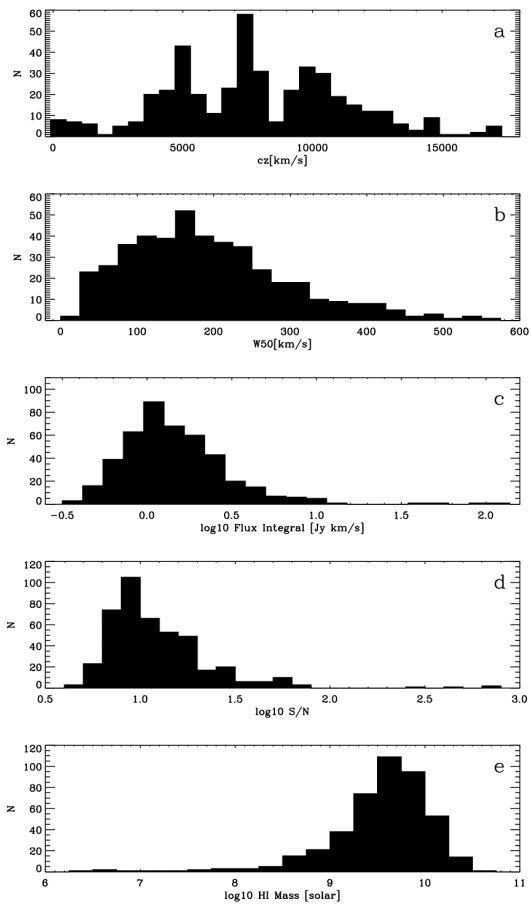}
\caption{Histograms of the HI detections with code 1 and 2 (i.e. excluding High Velocity Clouds): (a) heliocentric recessional velocity in \kms, (b) HI line width at half power ($W50$) in \kms, (c) logarithm of the flux integral in Jy \kms, (d) logarithm of the signal-to-noise ratio, and (e) logarithm of the HI mass in solar units.  \label{hists}}
\end{figure}

In Figure \ref{hists}, we show a series of histograms representing the heliocentric velocity, velocity width, flux integral, $S/N$ and HI mass distributions of all the detections with code 1 or 2 (i.e. the HVCs are excluded). The velocity distribution (Fig.\ref{hists}a) is dominated by large scale structure and the drop of sensitivity at higher velocities.  The median redshift of the distribution is 7550 \kms, which is just beyond the ridge of the Pisces-Perseus Supercluster. The galaxies in this catalog have velocity widths, $W50$, that range from 20 \kms \ to 550 \kms (see Fig.\ref{hists}b). Many of the very narrow signals come from nearby galaxies, which will turn out to be the lowest mass galaxies in the sample. The cutoff at 550 \kms \ is similar to the upper-limit of the velocity width distribution observed in previous large HI samples of galaxies \citep[e.g.][]{koribalski04,springob05sfi,spekkens06}. None of the profiles of the broadest galaxies with $W50>450$ \kms \ appear to be contaminated by neighboring galaxies and artificially broadened.

The third panel of Figure \ref{hists} shows the integrated flux distribution. The distribution of fluxes starts at a value of about 0.35 Jy \kms, which puts galaxies with very narrow HI lines just above the signal-to-noise threshold for inclusion in this catalog without corroborating observations. The median of the distribution is 1.3 Jy \kms, which is smaller than the faint-end cutoff of the integrated fluxes distribution of the galaxies in the HIPASS sample ($\sim1.5$ Jy \kms, see Fig.9 of \citet{meyer04}). The galaxies that stand out of the distribution with large integrated fluxes in excess of 40 Jy \kms \ are the nearby galaxies UGC1249 (IC1727), 1256 (NGC672, see next section), 12732 and 12754, which are all located within about 10 Mpc.

\begin{figure}[h]
\plotone{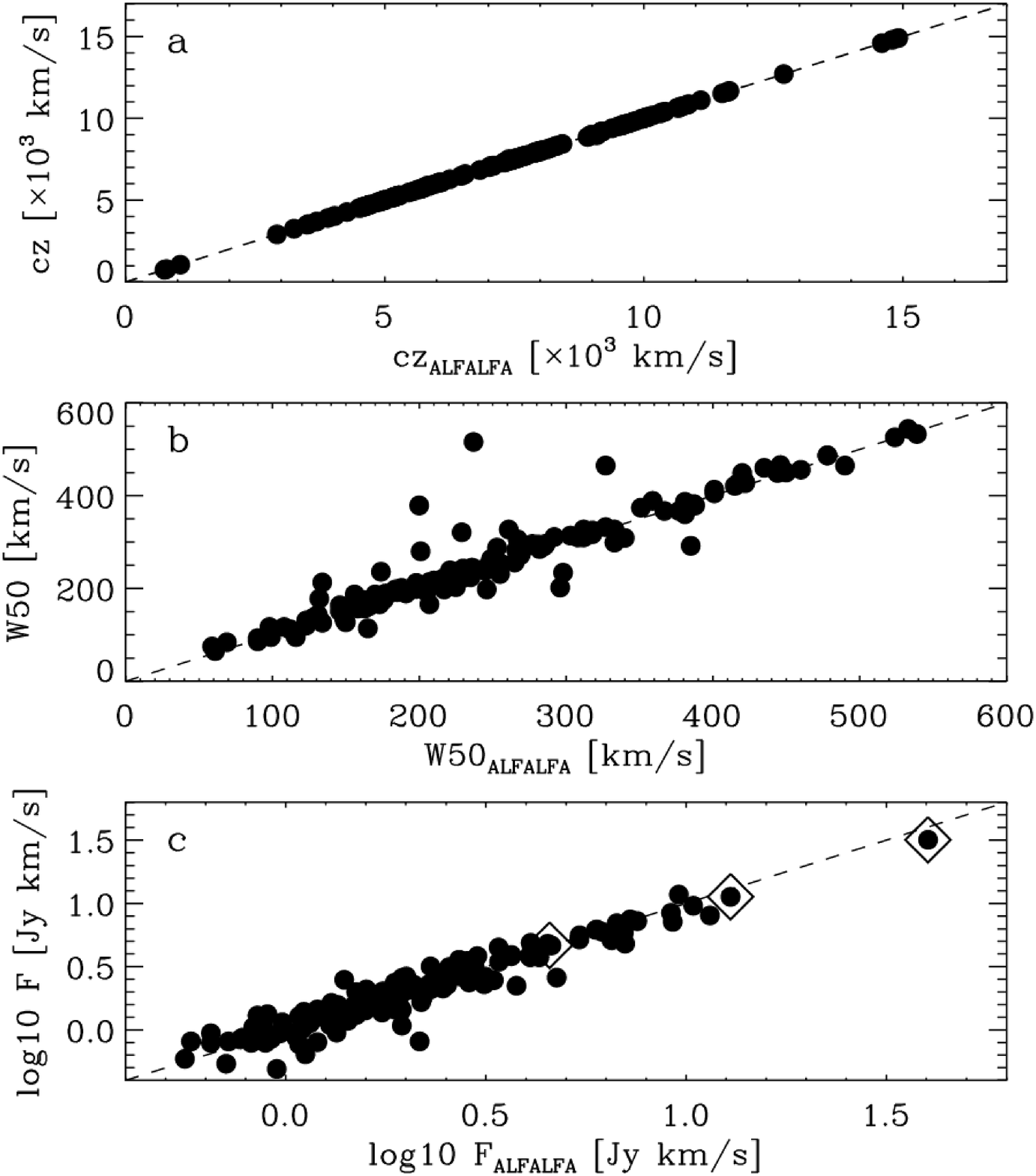}
\caption{Comparison between the properties of galaxies as measured by ALFALFA and by the pointed HI observations compiled in \citet{springob05sfi}: (a) heliocentric recessional velocity in \kms, (b) HI line width at half power in \kms, (c) logarithm of the flux integral in Jy \kms. The diamonds in panel (c) identify the three galaxies with optical diameters in excess of 2\arcmin.  In each case, the dashed line represents the 1:1 relation. \label{comp_archive}}
\end{figure}

We compare these values of redshifts, widths and fluxes to those measured through pointed HI observations. The digital HI archive of \citet{springob05sfi}, a homogeneous compilation of HI spectra for $\sim8800$ galaxies, is used as the reference. There are 141 of the galaxies in this ALFALFA catalog that are also found in the HI archive. The comparison between the main parameters of these galaxies is made in Figure \ref{comp_archive}. In all three cases, there is no systematic offset between the two sets of measurements. In the bottom panel, the three galaxies with optical diameters in excess of 2\arcmin \ are shown: assuming that the HI disk of a galaxy is 1.5 times larger than the optical disk, these three galaxies are resolved by the Arecibo beam at 21 cm.  As mentioned earlier, even though the method used to measure fluxes for the ALFALFA galaxies is optimized for sources with extents smaller than the beam, the flux of these three galaxies is adequately recovered.

As shown in the last panel (Fig.\ref{hists}e), we find 11 galaxies with $\log_{10}(M_{HI})<8.0$. Interestingly, all 4 galaxies with $\log_{10}(M_{HI})<7.0$ (and 5 of the 6 with $\log_{10}(M_{HI})<7.5$) are found in the vicinity of the nearby galaxy pair NGC672/IC1727 and their neighboring dwarf irregular galaxy NGC 784 (see \S \ref{ngc672} for details).  The median of the mass distribution is $4.3\times10^9$ \msun, while the galaxy with the largest HI mass in this catalog is UGC 12193 with $4.1\times10^{10}$ \msun, which puts it at the massive end of the HI mass function but is not atypical of galaxies with a morphological type such as this one \citep{springob05}.

\begin{figure}[ht!]
\plotone{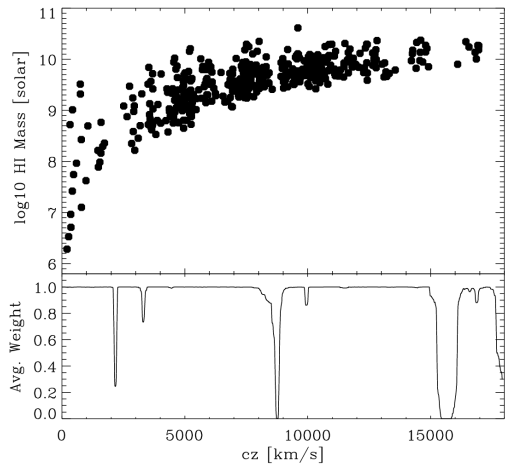}
\caption{{\it Top panel:} Distribution of measured HI mass as a function of heliocentric recession velocity. {\it Bottom panel:} Average weight of the ALFALFA data cubes as a function of velocity. \label{HImass}}
\end{figure}

The distribution of the HI mass of the detections as a function of heliocentric recession velocity is presented in Figure \ref{HImass}. The bottom panel gives an indication of the quality of the survey coverage at any given distance. This quality indicator is the weight factor of all the grid points as recorded when the data cubes were assembled. A weight of 1 indicates perfect survey coverage while a lower value means that part or all of the data were lost to rfi or incomplete coverage. The large gap in survey coverage at distances of $\sim225$Mpc ($cz\sim15500$\kms) corresponds again to airport radar rfi. The other gaps at 125 Mpc ($cz\sim8800$\kms), 47 Mpc ($cz\sim3300$\kms) and 31 Mpc ($cz\sim2200$\kms) are harmonics of the main radar frequency. Any other underdensities in the distribution of galaxies are caused by large scale structure, for example the large gap between 1500 and 2500 \kms \ which cannot entirely be explained by the presence of rfi.

\subsection{The NGC672/IC1727 and NGC784 Groups \label{ngc672}}

\begin{figure}[h]
\plotone{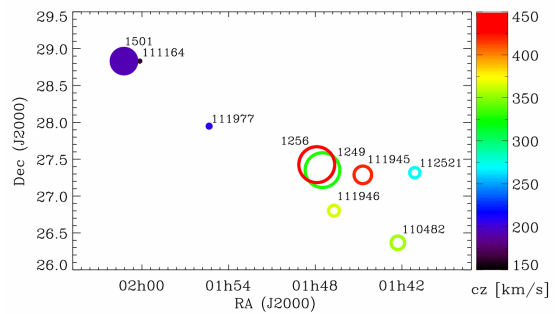}
\caption{Distribution on the sky of the galaxies in the NGC672 group (open symbols) and the NGC784 group (filled symbols) detected by ALFALFA. The AGC numbers of the galaxies (as seen in Tab.\ref{ngc672tab}) are labeled, and the size of the symbols are proportional to the logarithm of the ratio between the masses of the galaxies and the least massive galaxy of the sample.  The different symbols are color-coded by redshift (a color figure is available in the electronic version of the journal). \label{groups_plot}}
\end{figure}

The NGC672/IC1727 galaxy pair is a prominent feature of the local Universe in the sky area studied here. From the luminosity of the brightest red supergiant stars, \citet{sohn96} assign a distance of $7.9^{+1.0}_{-0.9}$ Mpc to NGC672, while \citet{karach04} report a distance of 7.2 Mpc for both NGC672 and IC1727 based on the luminosity of the brightest star. \citet{hucht00} assign three nearby dwarf irregular galaxies to membership in the group. These three satellites are detected by ALFALFA, as well as a fourth one, whose discovery was first reported in \citet{alfalfa2}.  The positions on the sky of these galaxies are illustrated in Figure \ref{groups_plot} as open symbols which are color-coded by redshift. In Table \ref{ngc672tab} we summarize the parameters of all the members of the NGC672 group. The first column is the AGC number and the second column an alternate commonly used name. Columns 3, 4 and 5 summarize information contained in Table \ref{params} concerning heliocentric recession velocity, velocity width and integrated flux. In column 6, we give the distance calculated using the peculiar velocity model, while the numbers in parenthesis are distances found in the literature and assigned based on primary distance indicators or group membership \citep{hucht00,karach03,karach04}. Finally, column 7 gives two estimates of the HI mass, based on each of the distance values, and column 8 the group membership. We note that the distances produced by the peculiar velocity model are inferior to the primary distances, which leads to an underestimate of the HI masses. Since the primary distances should be trusted over the peculiar velocity-inferred ones, this second mass estimate is adopted.   The total mass in HI contained in the six galaxies of the NGC672 group detected by ALFALFA is $2.5\times10^9$ \msun, of which 97\% is contained in the two large members of the group, NGC672 and IC1727. The NGC672 group is briefly considered by \citet{karach05} in a study of neighboring galaxy groups, even though no mass estimate is given on the grounds that accurate distances for individual members of groups located within 5 and 10 Mpc are not of satisfactory quality. 

There is a second group, or rather association of dwarfs, in the vicinity of the NGC672 group. The main member of this group is 
the dwarf galaxy NGC784, which is estimated to be at a distance of 5.0 Mpc, based on the luminosity of the brightest blue stars \citep{droz00}. The dwarf AGC111977 is located $1.9^{\circ}$ to the East of NGC672 and $1.7^{\circ}$ to the South-West of NGC784 (see Figure \ref{groups_plot}). With a recession velocity of 207 \kms \ and a tip of the red giant branch distance of 4.7 Mpc \citep{karach03}, it is identified as a member of the NGC784 group \citep{hucht00}. In Table \ref{ngc672tab} we give the parameters for three confirmed members of the NGC784 group, including two which are not found in Table \ref{params} because their declination is outside of the range studied here. They are NGC784 itself, and its satellite companion AGC111164.  These three members of the NGC784 group are plotted as filled symbols in Figure \ref{groups_plot}.  In their study of nearby dwarf galaxies associations, \citet{tully06} identify another galaxy with the NGC784 group, UGC1281 which is located 4.5 degrees (which is $\sim400$ kpc at the distance of the group) away from NGC784 and far out of the declination strip surveyed here. Using information on distances and redshifts for these four members, they derive a virial mass of $1.7\times10^{11}$ \msun \ for the NGC784 group, but not without saying that such a group is unlikely to be in dynamical equilibrium and therefore shedding some doubt on the validity of any mass calculations for these small associations.  Furthermore, as seen in Figure \ref{groups_plot}, the groups have very few members, and these few members do not seem to form relaxed structures. \citet{tully88} identifies galaxies of the NGC672 and NGC784 groups to the Triangulum Spur, suggesting that they may all be part of a larger structure.  Just as it did for the very low mass, low surface brightness dwarf AGC112521, we are hoping that ALFALFA will detect additional members of these groups, outside of the sky area covered here, to help us trace the relation between the different groups in this region of space.

As an example of that, we note that we also detect in the vicinity of NGC784 several HI clouds with recession velocity in the range of 30 to 50 
\kms. Just like NGC784 and AGC111164, these detections are made outside of the declination range covered in this data release and therefore are not included in Table \ref{params}. The two most interesting candidate extragalactic objects are however presented as the last two items in Table \ref{ngc672tab}. The first detection is made only 1.1\arcmin \ to the North of the 2MASS galaxy J02034519+2910528, but due to the relatively large size of the offset (see Fig. \ref{pointing}) and the significant spatial extent of the HI emission, the association is most likely coincidental. The other candidate detection has no apparent optical counterpart. If at the distance of NGC784, these two objects would have HI masses consistent with those of the known members of the group. Follow-up observations to come will try to corroborate the association of these two systems with the NGC784 group, even though the relatively large velocity offset between these two clouds and the other members of the group as well as the known population of Galactic clouds with similar velocities seem to argue against such an association.

\begin{deluxetable*}{cccccccccc}
\tablewidth{0pt}
\tabletypesize{\footnotesize}
\tablecaption{Detections in the NGC672 and NGC784 Groups \label{ngc672tab}}
\tablehead{
\colhead{AGC}   &\colhead{Other Name} & \colhead{HI Coords (2000)} & 
\colhead{cz$_\odot$}  & \colhead{$w50$} &
\colhead{$F_{c}$} & \colhead{Dist}  & \colhead{$\log M_{HI}$} & \colhead{Group} 
    \\
 & & & {\kms} & {\kms} & {Jy \kms} & Mpc & {$M_\odot$} & &
}
\startdata
112521 &\nodata & 014106.4+271903 &    274 &    26 &   0.65 &  4.6 (7.2) &   6.50 (6.89) &  N672\\
110482 &KK13    & 014216.8+262158 &    357 &    30 &   1.24 &  5.6 (7.2) &   6.96 (7.18) &  N672\\
111945 &KK14    & 014442.4+271717 &    423 &    36 &   2.80 &  6.4 (7.2) &   7.43 (7.53) &  N672\\
111946 &KK15    & 014642.3+264806 &    367 &    21 &   0.72 &  5.7 (7.2) &   6.74 (6.94) &  N672\\
1249   &IC1727  & 014730.0+272107 &    330 &   115 &  88.51 &  5.2 (7.2) &   8.76 (9.03) &  N672\\
1256   &NGC672  & 014753.9+272535 &    429 &   205 & 111.02 &  6.5 (7.2) &   9.04 (9.13) &  N672\\
111977 &KK16    & 015521.1+275656 &    207 &    29 &   0.78 &  3.7 (4.7) &   6.39 (6.61) &  N784\\
111164 &KK17    & 020009.3+284954 &    164 &    27 &   0.54 &  3.1 (4.7) &   6.10 (6.45) &  N784\\
1501   &NGC784  & 020115.0+284953 &    193 &    88 &  53.10 &  3.5 (5.0) &   8.18 (8.49) &  N784\\
122834 &\nodata & 020347.0+291153 &     51 &    10 &   0.85 &  \nodata \tablenotemark{a} &   (6.69)\tablenotemark{b} & \nodata \\
122835 &\nodata & 020533.0+291358 &     30 &    23 &   0.75 &  \nodata \tablenotemark{a} &   (6.64)\tablenotemark{b} & \nodata \\
\enddata
\tablenotetext{a}{No association between the HI detection and an optical galaxy which is part of the NGC784 group has been made. This object is likely to 
be of Galactic origin, and therefore a distance from the peculiar velocity model is not warranted.}
\tablenotetext{b}{HI mass of the object, if at the distance of NGC784 of 5.0 Mpc.}
\end{deluxetable*}

\subsection{High Velocity Clouds \label{hvc}}

The sky area covered in this study corresponds to a region of space known to be largely filled with high velocity gas, especially at large negative recession velocities. The most prominent structures in this overall region are Wright's Cloud \citep{wright79}, and filaments associated with a northern extension of the Magellanic Stream first detected by \citet[][thereafter BT04]{bt04}. This last study provided a first sensitive map of HVCs in this part of the sky. We show here how ALFALFA achieves even better sensitivity and will therefore improve significantly the completeness of HVC catalogs over the sky area it surveys.

In Table \ref{params} we give positions, velocities, widths and fluxes for \nhvc \ HVCs, some of which are isolated compact clouds and some of which are bright regions of more extended features. A note of caution needs to be made concerning the measurements made on the HVCs however. We used the same software to calculate fluxes for the HVCs as we did for all the extragalactic sources. This technique is optimized for compact sources with sizes comparable or smaller than the size of the beam, and will therefore not provide accurate fluxes for the large HVCs with extents comparable to or greater than the size of the ALFALFA data cubes. In these cases, we measured the fluxes of the bright knots in the clouds, which had $S/N$ above our usual detection threshold, and report these in Table \ref{params} with a note to that effect. These should be taken as indication of position and relative brightness of the different clouds, but a separate analysis, which is out of the scope of this paper, will be presented to provide accurate fluxes, brightness temperatures and other structural parameters of HVCs.

\begin{figure}[h]
\plotone{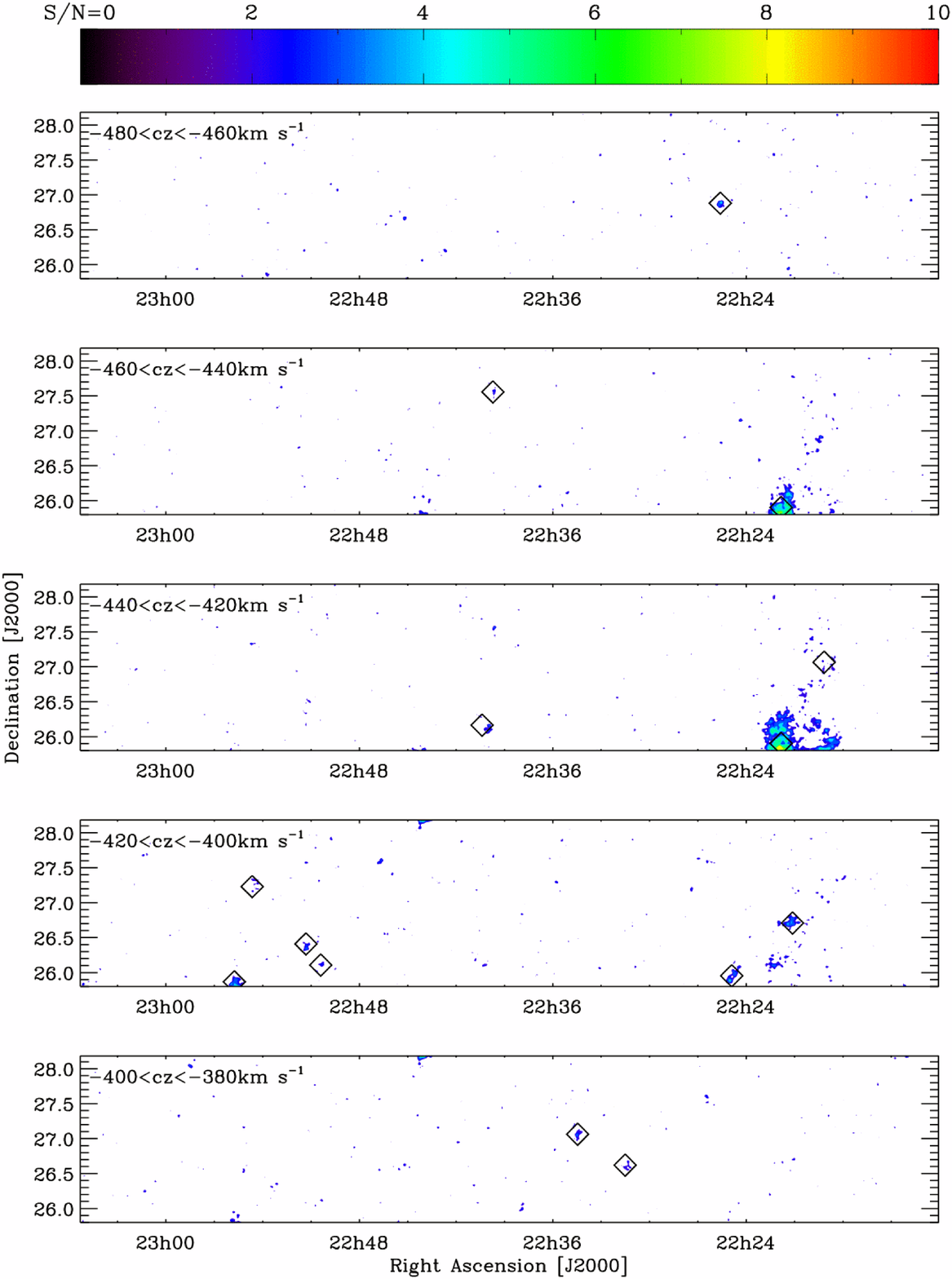}
\caption{Distribution of HVCs in the 22.2-23.1h range (region A in Fig.\ref{sky}), in five different cz intervals, represented as filled contours. The diamonds show the position of the knots measured and listed as HVCs in Table \ref{params}.  A color version of this figure is available in the electronic version of the journal.\label{hvc22h}}
\end{figure}

\begin{figure}[h]
\plotone{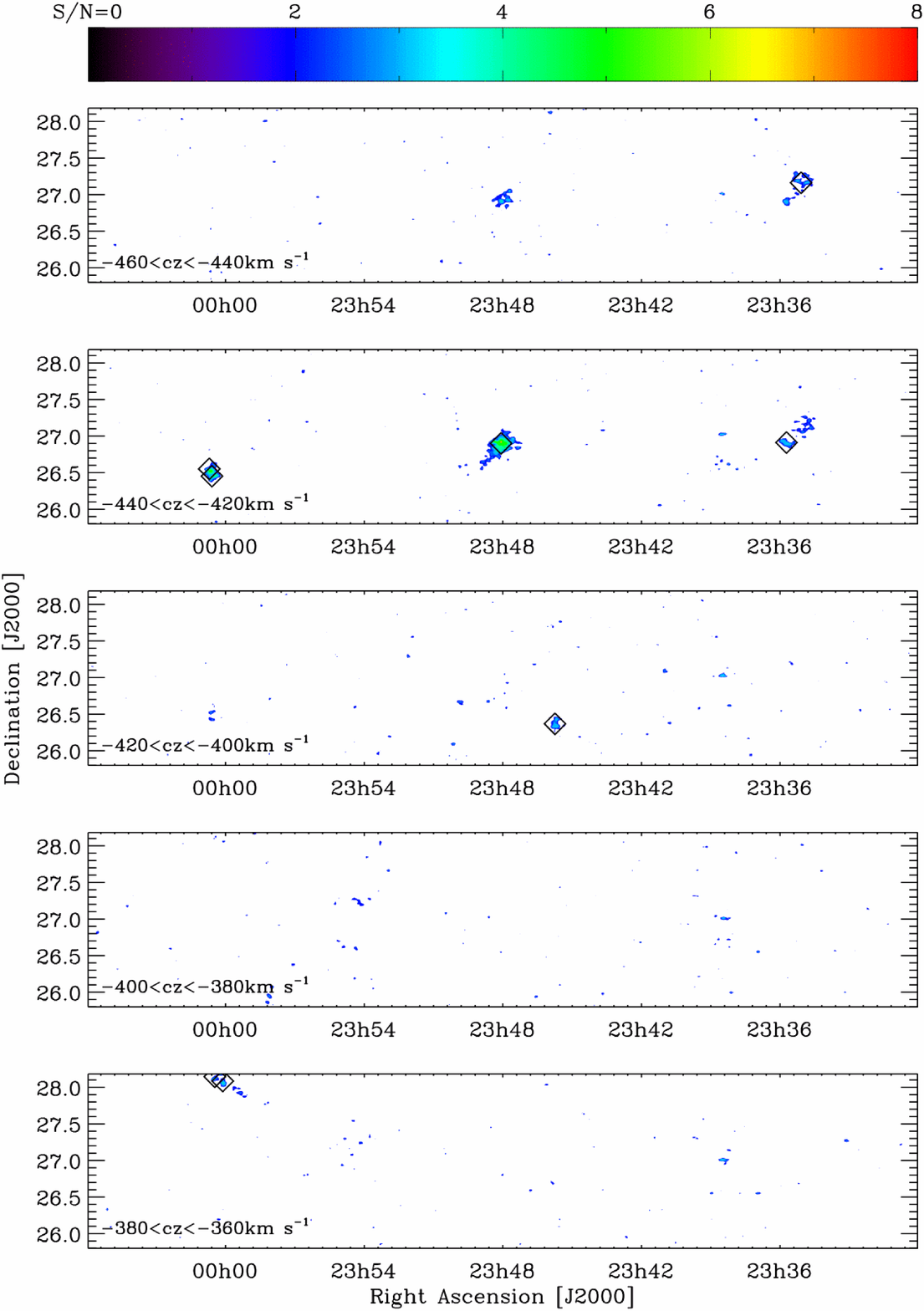}
\caption{Same as Figure \ref{hvc22h}, but for the 23.5-0.1h range (region B in Fig.\ref{sky}).  A color version of this figure is available in the electronic version of the journal.
\label{hvc23h}}
\end{figure}

\begin{figure}[h]
\plotone{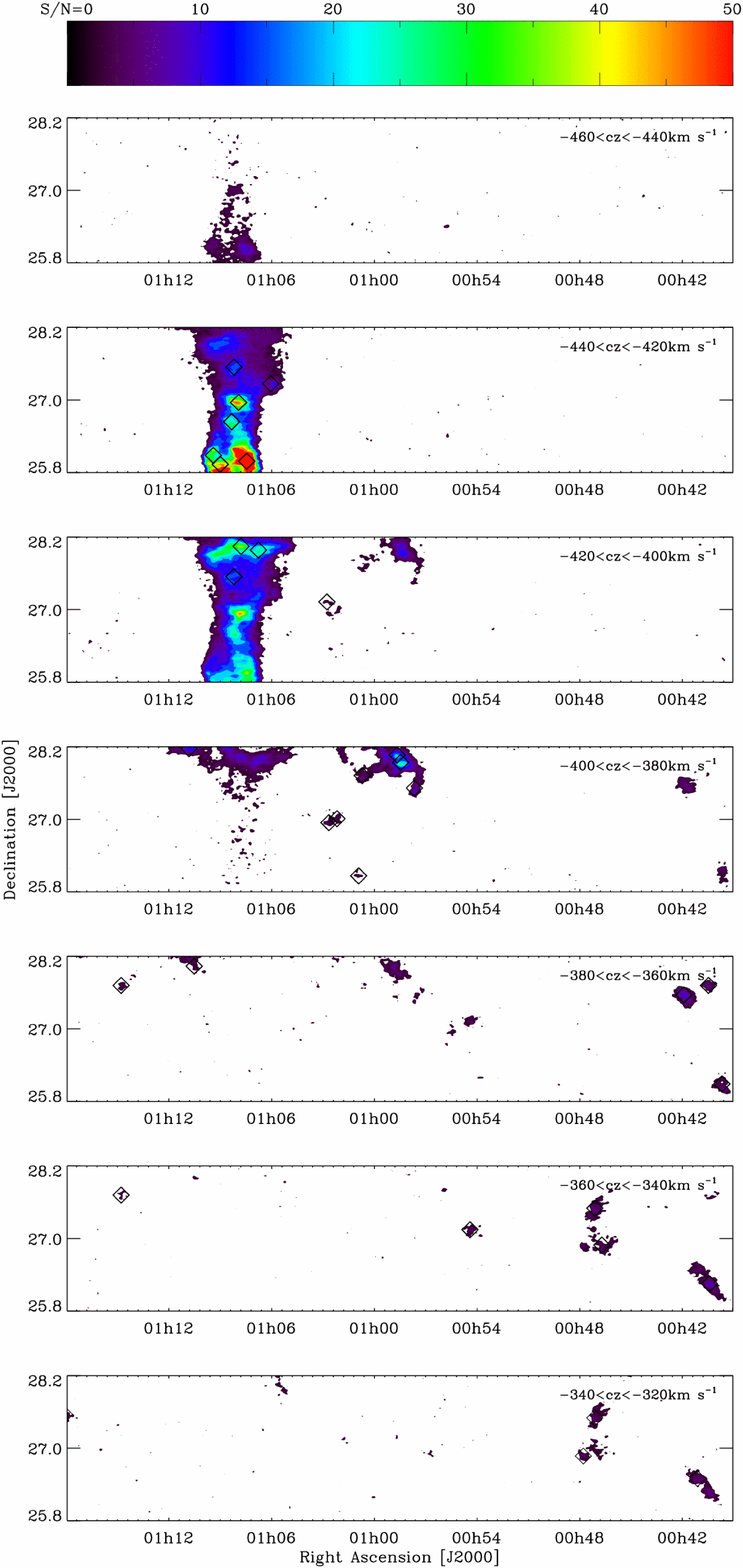}
\caption{Same as Figure \ref{hvc22h}, but for the 0.65-1.3h range (region C in Fig.\ref{sky}).  A color version of this figure is available in the electronic version of the journal.
\label{hvc01h}}
\end{figure}

In the bottom panel of Figure \ref{sky}, where we have shown the distribution of the HVCs, we have also defined three regions A, B and C which contain most of the interesting HVC features. In Figures \ref{hvc22h}, \ref{hvc23h} and \ref{hvc01h}, we present contour plots of the distribution of the HI for these three sky regions, respectively (note the different color scale in the different figures). In each of the figures, an individual panel represents a cut through the data cubes that is 20 \kms \ wide.  The diamonds indicate the position of the knots that have been measured and cataloged in Table \ref{params}. Regions A and B correspond to filaments identified by BT04 to be a northern extension of the Magellanic Stream.

The HVCs found in the right ascension range from $22^{\rm h}00^{\rm m}$ to $23^{\rm h}00^{\rm m}$ (Fig. \ref{hvc22h}) are mostly faint, compact and scattered, except for a brighter, very extended feature that appears centered on $22{\rm h}22^{\rm m}$ and $+26^{\circ}$ and extends North about 1.5$^{\circ}$. A future data release will contain the data South of $+26^{\circ}$ and reveal the southern extent of the cloud. Figure \ref{hvc23h} shows the HVCs observed between $23^{\rm h}30^{\rm m}$ and $00^{\rm h}00^{\rm m}$ and having $-460$ \kms $<cz_{\odot}<-360$ \kms. These again are compact and not especially clustered. The bright feature seen at around $01^{\rm h}08^{\rm m}$ (see Fig. \ref{hvc01h}) is the southern extension of Wright's cloud \citep{wright79}, which projects in the vicinity of the Local Group galaxy M33. Its velocity gradient from the North to South is very significant, with features from about -350 \kms down to -475 \kms. Several bright clouds are also seen at smaller velocities. Wright's cloud in itself is a significant feature because of its very high velocity, and its possible connection with M33 \citep{wright79}. A farther ALFALFA data release will contain the data for its main body as well as for HI emission around M33 and will shed light on the possible connection between the two systems, since the ALFALFA maps have higher spatial resolution and sensitivity than the Westerbork maps of both BT04 and \citet{westmeier05}. Interesting maps will also come from a separate project (``TOGS'', M.Putman P.I.) conducted commensally with ALFALFA using the GALSPECT spectrometer \citep[see][for a description of similar observations]{peek07}.  These observations will provide maps with the same spatial resolution, but with a spectral resolution of 0.2 \kms \ over the velocity range of $\pm 700$ \kms.

\begin{figure}[h]
\plotone{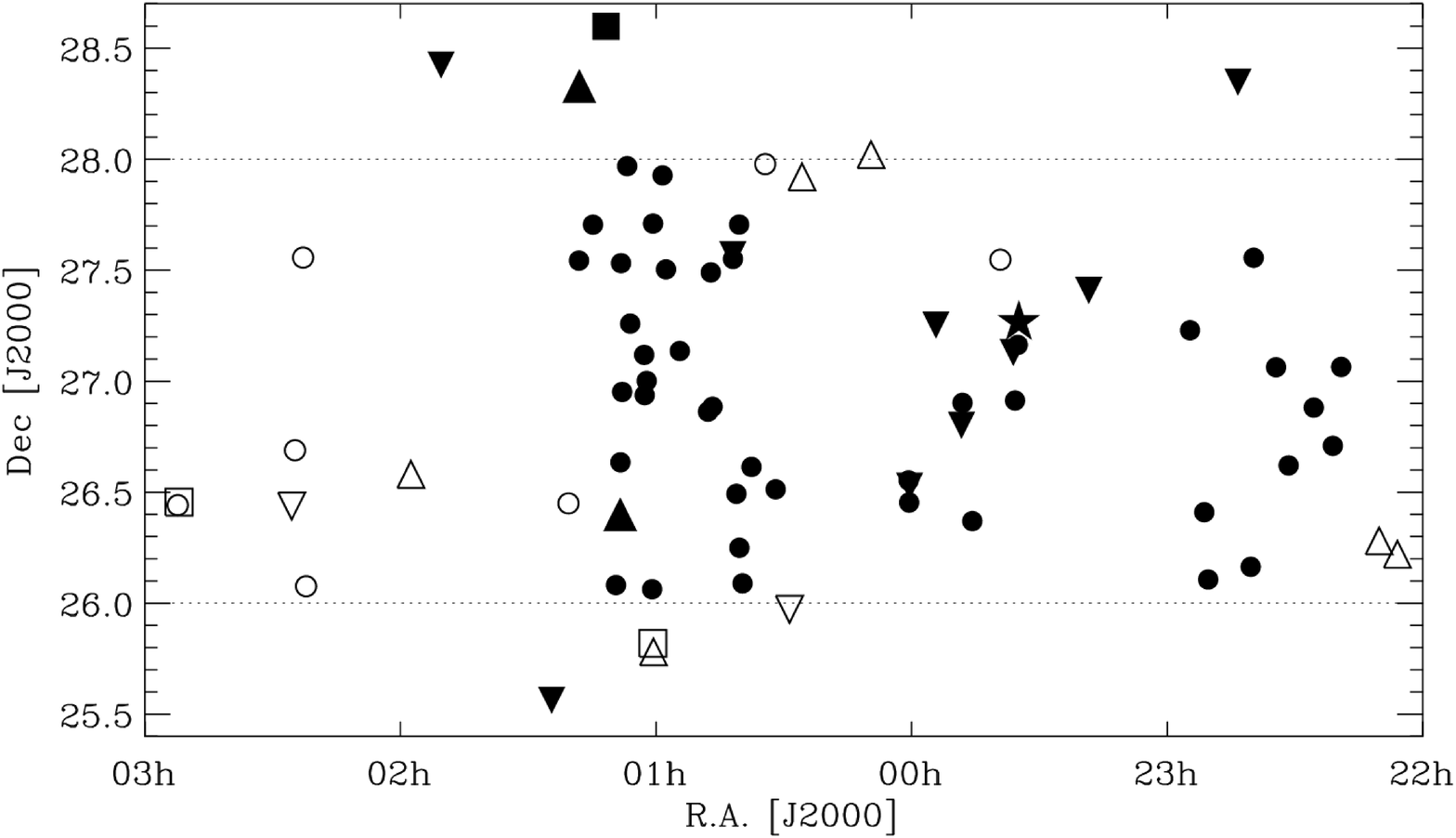}
\caption{Distribution on the sky of the HVCs detected by ALFALFA compared with previous surveys. The circles are the data from this study, the squares from \citet{wakker91}, the triangles from \citet{deheij}, the star is from the survey of \citet{lockman} and the upside-down triangles from \citet{bt04}. The filled symbols are detections with redshifts more negative than $cz_{\odot}=-200$ \kms, while the open symbols have $\mid cz_{\odot}\mid<200$ \kms.  Only the central positions of the clouds are plotted, as their overall sizes vary in the different catalogs.  The two horizontal lines outline the declination range for which ALFALFA measurements are available at the moment.\label{poshvc}}
\end{figure}

We now compare all these detections to previous catalogs of HVCs. In Figure \ref{poshvc}, we show the distribution of all the HVCs found in Table \ref{params}, as well as those from previous surveys. The four surveys we choose as comparison are (1) the first Dwingeloo survey of \citet{hulsbosch} as presented by \citet[][WW91]{wakker91} which covered the northern sky on a $1^{\circ}\times1^{\circ}$ grid with a detection limit of 0.05K, (2) the Leiden/Dwingeloo HI Survey \citep[][DBB02]{deheij} which covered the sky north of $-30^{\circ}$ on a $0.5^{\circ}\times0.5^{\circ}$ with an average sensitivity of 0.07K, (3) the sensitive targeted survey of \citet[][L02]{lockman}, which looked for faint HVCs with 860 pointings of the NRAO 140 Foot Telescope, a third of which were toward a bright extragalactic object, and produced a detection sensitivity of 14mK, and finally (4) a $60\times30$ degree coarse resolution survey conducted with the Westerbork Synthesis Radio Telescope (WSRT) to look for faint compact HVCs near M31 (BT04). For ease of comparison, in Figure \ref{poshvc} we distinguish between high negative velocity (filled symbols) and low velocity (open symbols) detections for each of the four catalogs (ALFALFA: circles, WW91: squares, DBB02: triangles, L02: star, BT04: upside-down triangles).

There are three detections reported by WW91 in and around the ALFALFA sky area analyzed here, and eight by DBB02, which are all summarized in Table \ref{hvctab}. The detections of both WW91 and DBB02 just above $+28^{\circ}$ and below $+26^{\circ}$ are confirmed by ALFALFA but not included in this data release. The northernmost feature is the main body of Wright's Cloud. The third cloud from WW91, WW432, is also found in the ALFALFA data (object 2-283 in Tab.\ref{params}). However, there is a significant discrepancy with the rest of the detections of DBB02. Their three reported detections with $v_{LSR}>0$ \kms \  (HVC293, HVC395 and HVC499 in the DBB02 catalog) are not detected by ALFALFA. In fact, ALFALFA makes no detections of high velocity clouds with positive recession velocity in this region of the sky. With the fluxes given by DBB02 all in excess of 30 Jy \kms, these sources should have easily been detected in our survey. The same is true for HVC286 reported by DBB02 to be at $v_{LSR}=-123$ \kms, though in this case given the poor spatial resolution of the survey of DBB02 it is not entirely impossible that ALFALFA would have detected the cloud if it's coverage had extended below 22h in RA. Finally, we observe HI emission consistent with HVC408, which seems part of a large filament that blends in with the main Galactic emission, and as such has not been measured and reported in Table \ref{params} as an HVC.

Only three of the lines of sight observed by L02 go through our surveyed sky area. They report one detection, which is confirmed by ALFALFA and measured as sources 2-414 and 2-417. These ultra compact HVCs are typical of the clouds detected in our survey, with dimensions of about $0.5^{\circ}$ or less. The ALFALFA $6\sigma$ HI column density limit for a cloud resolved by the beam, having a velocity width of 20 \kms \ and observed at a spectral resolution of 10 \kms \ is $3.2\times10^{18}$ atoms cm$^{-2}$, which is close to that achieved in the very high sensitivity, though small area survey of L02.

There is a much better agreement between our HVC detections and those of BT04, who report several detections over the sky area covered here. Our observed distribution of HVC corresponds to the large filaments and loops of HI reported in this study that they associate with the Magellanic Stream. We however detect a significant number of knots in the distribution of HI between 22h and 23h that are uncatalogued by BT04. These 10 ALFALFA detections have integrated fluxes in the range between 0.5 and 1.5 Jy \kms, while the faintest source in the catalog of BT04 has a flux of 2.8 Jy \kms. When ALFALFA is completed, it will therefore be possible to extend the detailed HVC survey of BT04 to a sky area 4 times as large and provide a much needed high sensitivity map of the high velocity HI emission.

\begin{deluxetable*}{cccccc}
\tablewidth{0pt}
\tabletypesize{\footnotesize}
\tablecaption{Detections of HVCs Previously Reported \label{hvctab}}
\tablehead{
\colhead{Catalog Name}   &\colhead{Source} & \colhead{HI Coords (2000)} & 
\colhead{$v_{LSR}$}  & \colhead{$F$} & \colhead{in ALFALFA?} 
    \\
 & & & {\kms} & {Jy \kms} & 
}
\startdata
HVC286	&DBB02	&2206.0+2613	&-123	&623	&no \\
HVC293	&DBB02	&2210.3+2617	&214	&61	&no \\
L119	&L02	&2334.8+2716	&-438	&\nodata &yes \\
HVC395	&DBB02	&0009.5+2801	&99	&56	&no \\
HVC408	&DBB02	&0025.7+2755	&-110	&175	&yes \\
HVC442	&DBB02	&0100.6+2547	&-125	&107	&yes \\
WW478	&WW91	&0100.7+2549	&-128	&156	&yes \\
HVC448	&DBB02	&0108.4+2624	&-429	&558	&yes \\
WW466	&WW91	&0111.7+2836	&-392	&1730	&yes \\
HVC459	&DBB02	&0118.0+2820	&-367	&54	&yes \\
HVC499	&DBB02	&0157.5+2635	&264	&30	&no \\
WW432	&WW91	&0251.9+2627	&-131	&18	&yes \\
\enddata
\end{deluxetable*}

\subsection{Distribution of Galaxies and the Local Void}

As mentioned earlier, the ALFALFA survey was designed to cover a broad range of cosmic environments in order to study the properties of HI-rich systems as a function of local environment. The catalog presented here gives us a first chance of looking at a large nearby void. With the high sensitivity of ALFALFA, we will be able to put strong constraints on the abundance of gas-rich dwarf galaxies in voids to ease the comparison with the predictions of theoretical models and numerical simulations. Due to its proximity, the Pisces-Perseus foreground void is an excellent environment to look for such dark galaxies. 

\begin{figure}[h]
\plotone{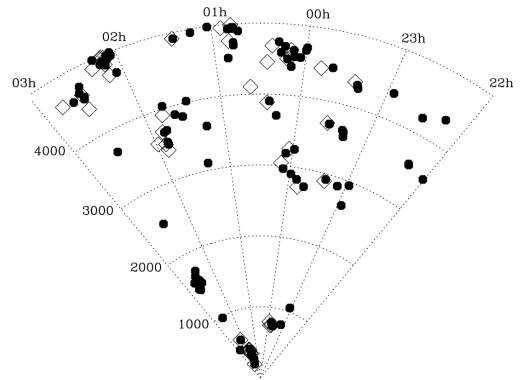}
\caption{Distribution of the sources with $cz_{\odot}<5000$ \kms. The circles are the ALFALFA detections, and the diamonds optical data from the AGC. \label{voidplot}}
\end{figure}

In Figure \ref{voidplot}, we show the distribution of all ALFALFA detections and galaxies with optical redshifts smaller than 5000 \kms. First of all, we notice how the distribution of the HI-selected sample closely follows the distribution of the galaxies optically detected. Secondly, we can clearly see a large void region both in the optical and ALFALFA data sets.  This region which roughly extends from $22^{\rm h}00^{\rm m}$ to $02^{\rm h}00^{\rm m}$ and from between $cz\sim1000$ to $cz\sim2500$ \kms \ contains no galaxies that could be detected either through the optical surveys or with ALFALFA. We used the density field from the IRAS PSCz redshift survey \citep{psczdens} to confirm the void nature of the selected volume of space. All the PSCz grid points within the volume are underdense compared to the cosmic mean ($\overline{\rho}$), with density contrasts in the range of $-0.8<\delta<-0.1$ and a median value of $\delta=-0.5$, where $\delta=(\rho-\overline{\rho})/\overline{\rho}$.

While the size of the dataset currently available does not allow for a reliable quantitative analysis of the situation, because only a fraction of one nearby void has been surveyed, we can already suspect that ALFALFA is not going to reveal a large population of dark galaxies in the voids. The fact that no galaxies were detected by ALFALFA in a large volume of 460 Mpc$^3$ of a nearby void, throughout which it is sensitive to objects with HI masses of $10^8$\msun \ is a good indication of that. For comparison, by scaling down the predictions for the large voids simulated by \citet{gott03} to the volume sampled by our current observations and assuming that $10\%$ of the mass of the dark halos is in the form of HI (which is admittedly probably a very enthusiastic estimate), ALFALFA should have detected 38 galaxies with M$_{HI}>10^8$\msun.

These results for the void phenomenon are also exacerbated by the fact that five of the six lowest mass objects in our sample are satellites of the nearby galaxies NGC672 and NGC784. Based on these results, there does not seem to be a tendency for low-mass systems to be more uniformly distributed than their more massive counterparts. There is also the fact that all HI detections with $cz_{\odot}>0$ have been matched with optical counterparts. Unless a number of the HVCs detected are extragalactic as discussed by several authors \citep[e.g.][]{oort,verschuur,giovanelli81,braun,blitz}, there is no evidence so far for dark galaxies either as satellites of massive halos or in underdense regions of space.

\section{Summary}
We present a catalog of \ntot \ HI detections from the ALFALFA survey over 135 deg$^{2}$ ($22{\rm h}<\alpha<03{\rm h}$,  $+26^{\circ}<\delta<+28^{\circ}$). Of these, \nhvc \ are identified as HVCs while the remaining \ngal \ are extragalactic and all associated with an optical counterpart. These matches are made with great confidence due to the relatively high pointing accuracy and spatial resolution of the survey. Of the extragalactic objects, $62\%$ are new detections made by ALFALFA and $59\%$ are new redshift measurements.

Five of the six objects detected with $\log_{10}(M_{HI})<7.5$ are satellites of either the galaxy pair NGC672/IC1727 or of their neighbor, the dwarf irregular galaxy NGC784. We find no evidence for low-mass optically-dark systems in the vicinity of these large nearby galaxies. We also do not find them in a 460 Mpc$^3$ slice through a nearby local void in the foreground of the Pisces-Perseus Supercluster.
The non-detection suggests that a large population of HI-rich yet optically-dark galaxies is unlikely to reconcile the discrepancies between the observed abundance of dwarf galaxies in voids and the predictions made on the basis of numerical simulations.

We have also used the ALFALFA data to study the HVC population. In addition to large bright clouds previously cataloged \citep{wakker91,deheij} such as the southern extension of Wright's Cloud, we detect a large number of ultra compact HVCs which correspond to the large filaments observed by \citet{bt04} and associated with a northern extension of the Magellanic Stream. When completed, ALFALFA and its commensal Galactic-ALFA survey will be the most sensitive large scale surveys for High Velocity Clouds, as evidenced by comparing the HVCs found in this catalog to previous detections.

While the data presented here are extracted from only $2\%$ of the full area ALFALFA will survey, we have been able to show the potential of the survey not only to significantly increase the number of galaxies for which HI redshifts are available, but also to better our understanding of the population of gas-rich galaxies, to put strong constraints on the abundance of HI-rich but optically-dark galaxies, and even to study the population of HVCs.

\acknowledgements
This work was supported by NSF grants AST-0607007, AST-0435697 and AST-0307661, and by grants from the {\it Fonds qu\'{e}becois de la recherche sur la nature et les technologies} and from the Brinson Foundation. We acknowledge the use of NASA's {\it SkyView} facility (http://skyview.gsfc.nasa.gov) located at NASA Goddard Space Flight Center. This research has also made use of the NASA/IPAC Extragalactic Database (NED) which is operated by the Jet Propulsion Laboratory, California Institute of Technology, under contract with the National Aeronautics and Space Administration.

\clearpage

\begin{landscape}
\begin{deluxetable*}{cccccccccccccc}
\tablewidth{0pt}
\tabletypesize{\tiny}
\tablecaption{HI Candidate Detections\tablenotemark{a}  \label{params}}
\tablehead{
\colhead{Source}  & \colhead{AGC}   & \colhead{$\alpha_{J2000}$(HI)} & \colhead{$\delta_{J2000}$(HI)} & \colhead{$\alpha_{J2000}$(Opt.)} & 
\colhead{$\delta_{J2000}$(Opt.)} & \colhead{cz$_\odot$}  & \colhead{$W50 ~(\epsilon_W$)} &
\colhead{$F_{c} ~(\epsilon_{f})$\tablenotemark{b}} & \colhead{S/N} & \colhead{rms} &
\colhead{Dist}    & \colhead{$\log M_{HI}$} & \colhead{Code}    \\
 & & & & & & {\kms} & {\kms} & {Jy \kms} & & {mJy} & Mpc & {$M_\odot$} &
}
\startdata
2-  1  & 331405 &      00 00 03.7 &       +26 00 56 &      00 00 03.5 &       +26 00 50 &  10409 &   316(  8) &   2.35(0.08) &   15.7 &   1.88 &  143.8 &  10.06 & 1   \\
2-  2  & 102571 &      00 00 17.2 &       +27 23 59 &      00 00 17.3 &       +27 24 03 &   4654 &   104(  3) &   1.83(0.06) &   19.0 &   2.10 &   65.9 &   9.27 & 1   \\
2-  3  &  12896 &      00 00 30.1 &       +26 19 28 &      00 00 31.4 &       +26 19 31 &   7653 &   170( 10) &   2.82(0.08) &   22.0 &   2.20 &  104.5 &   9.86 & 1   \\
2-  4  & 102576 &      00 00 35.3 &       +26 27 12 &                 &                 &   -430 &    21(  2) &   0.58(0.03) &   11.7 &   2.40 &        &        & 9 * \\
2-  5  & 102578 &      00 00 42.3 &       +26 33 11 &                 &                 &   -429 &    22(  3) &   0.64(0.04) &   12.8 &   2.32 &        &        & 9 * \\
2-  6  &  12920 &      00 02 23.1 &       +27 12 51 &      00 02 23.0 &       +27 12 38 &   7613 &   281(  2) &   2.60(0.09) &   16.0 &   2.17 &  103.9 &   9.82 & 1   \\
2-  7  & 102553 &      00 03 23.1 &       +27 24 29 &      00 03 25.1 &       +27 24 11 &   4544 &    83( 10) &   0.79(0.05) &    9.3 &   2.08 &   64.3 &   8.89 & 1   \\
2-  8  & 100025 &      00 04 43.9 &       +26 50 16 &      00 04 44.5 &       +26 49 57 &   7556 &   134( 15) &   0.98(0.06) &    9.8 &   1.93 &  103.1 &   9.39 & 1   \\
2-  9  & 102554 &      00 05 15.7 &       +27 29 19 &      00 05 15.3 &       +27 29 23 &  11854 &   232( 27) &   1.55(0.08) &   11.0 &   2.06 &  164.5 &  10.00 & 1   \\
2- 10  & 102555 &      00 05 17.4 &       +26 31 19 &      00 05 19.2 &       +26 31 10 &   7432 &   140( 33) &   0.68(0.06) &    7.0 &   1.85 &  101.3 &   9.22 & 1   \\
2- 11  &     40 &      00 05 49.2 &       +27 26 57 &      00 05 48.4 &       +27 26 56 &   7522 &   388(  2) &   2.28(0.10) &   12.8 &   2.02 &  102.6 &   9.75 & 1   \\
2- 12  & 100037 &      00 06 03.0 &       +27 20 52 &      00 06 02.5 &       +27 20 58 &   3182 &    63(  1) &   1.14(0.04) &   16.5 &   1.94 &   44.8 &   8.73 & 1   \\
2- 13  & 102556 &      00 06 07.5 &       +26 05 37 &      00 06 08.5 &       +26 06 44 &   7288 &   261( 67) &   1.19(0.09) &    8.1 &   2.03 &   99.3 &   9.44 & 1   \\
2- 14  & 102557 &      00 06 25.3 &       +27 55 00 &      00 06 25.4 &       +27 54 40 &   7527 &   270( 49) &   1.56(0.09) &   10.6 &   2.00 &  102.7 &   9.59 & 1   \\
2- 15  &     50 &      00 06 40.0 &       +26 09 09 &      00 06 40.1 &       +26 09 14 &   7556 &   401( 14) &   2.63(0.09) &   16.0 &   1.83 &  103.1 &   9.82 & 1   \\
\hline
\enddata
\tablenotetext{a}{The full contents of this table are presented in the online version of the journal, and are also available through our public digital archive at {\it http://arecibo.tc.cornell.edu/hiarchive/alfalfa/}. The description of the table entries can be found in \S \ref{cat} of this paper, and follows the format of the first ALFALFA catalog published in \citet{alfalfa3}.}
\tablenotetext{b}{For extended High Velocity Clouds, only the fluxes of the brightest knots of the clouds were in some cases measured.  See Column 14 and the appropriate notes for individual objects.}
\end{deluxetable*}
\clearpage
\end{landscape}

\end{document}